\documentclass[aps,nofootinbib,superscriptaddress, showpacs,preprintnumbers,  twocolumn, nofootinbibt]{revtex4-2}
\usepackage{eurosym}
\usepackage{amsmath}
\usepackage{bm}
\usepackage{amsfonts}
\usepackage{amssymb}
\usepackage{float}
\usepackage{graphicx}
\usepackage{subfig}
\usepackage{caption}
\usepackage[export]{adjustbox}
\graphicspath{ {folder} }

\def\be{\begin{equation}}
\def\ee{\end{equation}}
\def\bea{\begin{eqnarray}}
\def\eea{\end{eqnarray}}

\begin{document}

\title{Reply to Comment on  "Dark matter as a Weyl geometric effect"}
\author{Piyabut Burikham}
\email{piyabut@gmail.com}
\affiliation{High Energy Physics Theory Group, Department of Physics,
Faculty of Science, Chulalongkorn University, Bangkok 10330, Thailand,}
\author{Tiberiu Harko}
\email{tiberiu.harko@aira.astro.ro}
\affiliation{Department of Theoretical Physics, National Institute of Physics
and Nuclear Engineering (IFIN-HH), Bucharest, 077125 Romania,}
\affiliation{Department of Physics, Babes-Bolyai University, Kogalniceanu Street,
	Cluj-Napoca 400084, Romania,}
\affiliation{Astronomical Observatory, 19 Ciresilor Street,
	Cluj-Napoca 400487, Romania,}
\author{Kulapant Pimsamarn}
\email{fsciklpp@ku.ac.th}
\affiliation{Department of Physics, Faculty of Science, Kasetsart University, Bangkok 10900, Thailand}
\author{Shahab Shahidi}
\email{s.shahidi@du.ac.ir}
\affiliation{School of Physics, Damghan University, Damghan, Iran}

\date{\today }

\date{\today }

\begin{abstract}
In a recent Comment on the paper "Dark matter as a Weyl geometric effect", by Burikham et al., Phys. Rev. D 107, 064008 (2023), posted on arxiv. org as eprint arXiv:2306.11926, it was claimed that the exact solution found in the above mentioned paper by Burikham et al.  "is wrong". In this Reply to the Comment we present, in a clear and comprehensive way,  a step by step derivation of the exact solution of the vacuum static spherically symmetric field equations of the Weyl geometric gravity theory, and we show that, contrary to the claims in arXiv:2306.11926, the obtained solution is correct, and it satisfies all the equations of motion of the basic theory. Hence, it can be considered as a viable alternative model for the explanation of the behavior of the galactic rotation  curves, without invoking the presence of dark matter.

\end{abstract}


\maketitle

\tableofcontents

\section{Introduction}

In a recent comment \cite{1}  on the paper \cite{2}, it was claimed that the exact static spherically symmetric solution obtained in \cite{2}. as well as in \cite{3}, is incorrect mathematically, since it does not satisfy the field equations of the model. In the present Reply to the Comment \cite{1} we will not address the specific points raised in the analysis and considerations of the author of \cite{1}, but we will present again the full derivation of the vacuum solution of the field equations of Weyl geometric gravity, which will make clear that the solution presented  in \cite{2} is correct, and thus it can be used to investigate various astrophysical aspects, including the possibility of the description of the galactic dynamics. Our presentation will also invalidate the main claims and conclusions of \cite{1}.

Weyl geometry is an extension of the Riemannian geometry, with the Weyl connection given by \cite{2,3}
\begin{align}\label{connex}
& \tilde{\Gamma}^\lambda_{\mu\nu} = \Gamma^\lambda_{\mu\nu} + \frac{1}{2}
\alpha \Big[ \delta^\lambda_\mu \omega_\nu + \delta^\lambda_\nu \omega_\mu -
g_{\mu\nu} \omega^\lambda \Big],
\end{align}
where $\Gamma^\lambda_{\mu\nu}$ is the standard Levi-Civita connection, and $\alpha$ is a constant. The Weyl geometric gravity theory is the linearized version of the conformally invariant action, defined in a Weyl geometric framework, and which is given by \cite{Weyl1, Weyl2, 4,5,6}
\bea\label{inA}
S_0=\int{\Big[\, \frac{1}{4!}\,\frac{1}{\xi^2}\,\tilde R^2  - \frac14\, F_{\mu\nu}^{\,2} \Big]\sqrt{-g}d^4x},
\eea
where $\xi$ is a coupling constant, satisfying the condition  $\xi < 1$. For the definitions of the Weyl scalar $\tilde{R}$, of the strength of the Weyl vector $F_{\mu \nu}$, and of the geometric quantities in Weyl geometry see \cite{2}. The action (\ref{inA}) can be linearized in the Weyl scalar by introducing an auxiliary scalar field $\phi_0$ in $S_0$, and  by representing the term  $\tilde{R}^2$ as
\be
\tilde{R}^2\rightarrow -2 \phi_0^2\,\tilde R-\phi_0^4,
\ee

Hence,  we obtain the equivalent Weyl geometric action given by
\bea\label{alt3}
S_1=\int{\Big[-\frac{1}{12}\frac{1}{\xi^2}\,\phi_0^2\,\tilde R
-\frac14 \,F_{\mu\nu}^2-\frac{\phi_0^4}{4!\,\xi^2}
\Big]\sqrt{-g}d^4x} ,
\eea
which has the important property  of being linear in the Weyl curvature $\tilde{R}$. The gravitational action (\ref{alt3}) fully implements the Weyl gauge symmetry, and the conformal invariance.  Moreover, $S_1$ spontaneously
breaks down to an Einstein-Proca type action for the Weyl gauge field \cite{4,5,6}.

We substitute now in Eq.~(\ref{alt3}) the Weyl geomeyric quantities with their Riemannian counterparts, and after we redefine the physical and geometrical variables by means of a gauge transformation, we obtain the action of the Weyl geometric gravity as given by \cite{2, 4,5,6}
\begin{eqnarray}\label{a3}
S &=& \int d^4x \sqrt{-g} \Bigg[ -\frac{1}{12} \frac{\phi^2}{\xi^2}
\Big( R - 3\alpha\nabla_\mu \omega^\mu - \frac{3}{2} \alpha^2 \omega_\mu
\omega^\mu \Big) \nonumber\\
&&- \frac{1}{4!}\frac{\phi^4}{\xi^2} - \frac{1}{4} F_{\mu\nu}
F^{\mu\nu} \Bigg].
\end{eqnarray}

In order to obtain  the complete set of field equations corresponding to the action (\ref{a3}) we must vary it with respect to the metric tensor, the scalar field, and the Weyl vector, respectively. We vary first the action Eq.~(\ref{a3}) with respect to the metric tensor, and thus we obtain the field
equations of the Weyl geometric gravity theory as \cite{2}
\begin{eqnarray}\label{b2a}
&&\frac{\phi ^{2}}{\xi ^{2}}\Big(R_{\mu \nu }-\frac{1}{2}Rg_{\mu \nu }%
\Big)\nonumber\\
&&-\frac{3\alpha }{2\xi ^{2}}\Big(\omega ^{\rho }\nabla _{\rho }\phi
^{2}g_{\mu \nu }-\omega _{\nu }\nabla _{\mu }\phi ^{2}-\omega _{\mu }\nabla
_{\nu }\phi ^{2}\Big)\nonumber\\
&&+\frac{3\alpha ^{2}}{4\xi ^{2}}\phi ^{2}\Big(\omega
_{\rho }\omega ^{\rho }g_{\mu \nu }-2\omega _{\mu }\omega _{\nu }\Big)
+6F_{\rho \mu }F_{\sigma \nu }g^{\rho \sigma }-\frac{3}{2}F_{\rho \sigma
}^{2}g_{\mu \nu }\nonumber\\
&&-\frac{1}{4\xi ^{2}}\phi ^{4}g_{\mu \nu }
+\frac{1}{\xi ^{2}}%
\Big(g_{\mu \nu }\Box -\nabla _{\mu }\nabla _{\nu }\Big)\phi ^{2}=0.
\end{eqnarray}

As a second step we vary the action  Eq.~(\ref{a3})
with respect to the scalar field $\phi $. This variation gives the equation of motion of $%
\Phi $,
\begin{equation}\label{b4}
R-3\alpha \nabla _{\rho }\omega ^{\rho }-\frac{3}{2}\alpha ^{2}\omega _{\rho
}\omega ^{\rho }+\Phi =0.
\end{equation}%

Finally, we vary the action Eq.~(\ref{a3}) with respect to $\omega _{\mu }$,
thus obtaining the equation of motion of the Weyl vector as
\begin{equation}\label{Fmunu}
4\xi ^{2}\nabla _{\nu }F^{\mu \nu }-\alpha ^{2}\Phi\omega ^{\mu }+\alpha \nabla
^{\mu }\Phi=0.
\end{equation}%

The trace of Eq.~(\ref{b2a}) gives
\begin{equation}\label{b3n}
\Phi R+3\alpha \omega ^{\rho }\nabla _{\rho }\Phi +\Phi ^{2}-\frac{3}{2}%
\alpha ^{2}\Phi \omega _{\rho }\omega ^{\rho }-3\Box \Phi =0,
\end{equation}%
where we have denoted $\Phi \equiv \phi ^{2}$.

We multiply Eq.~(\ref{b4}) by $\Phi$, and we subtract the result  from Eq.~(\ref{b3n}). Thus we obtain the generalized Klein-Gordon equation for the scalar field $\Phi$ as given by
\begin{equation}\label{b5}
\Box \Phi-\alpha \nabla _{\rho }(\Phi\omega ^{\rho })=0.
\end{equation}%

In order to check the consistency of the model  we apply to both sides of Eq.~(\ref{Fmunu}) the operator  $\nabla _{\mu }$. Thus, we reobtain again Eq.~(\ref{b5}), a result that proves the mathematical consistency of the field equations of the Weyl geometric gravity  theory.

Our next goal is to find a vacuum, static, spherically symmetric solution of the full set of the field equations of the Weyl geometric theory.  We consider this problem in the next Section.

\section{An exact solution in Weyl geometric gravity}

We assume that the static, spherically symmetric vacuum of the Weyl geometric gravity can be described by the metric
\begin{equation}\label{line}
ds^{2}=e^{\nu (r)}dt^{2}-e^{\lambda (r)}dr^{2}-r^{2}\left(d\theta ^{2}+\sin ^{2}\theta d\varphi ^{2}\right),
\end{equation}
where the metric tensor components $\nu$ and $\lambda$ are functions of the radial coordinate $r$ only.

Due to the choice of our metric, and of the symmetry of the problem,  the third and the
fourth components of the Weyl vector  must vanish identically. Hence, in spherical symmetry, the Weyl vector has generally the components $\omega _{\mu }=\left( \omega _{0}(r),\omega_{1}(r),0,0\right) $.

However, in order to investigate the possibility of the existence of an exact solution, we introduce a further simplification of our formalism, by assuming that the temporal component, $\omega _{0}$, of the Weyl vector also vanishes identically, $\omega _0\equiv 0$. Therefore, we assume that the Weyl vector $\omega _{\mu}$ has only one non-zero radial component, and thus
\be
\omega _{\mu }=(0,\omega _{1}(r),0,0).
\ee

As a consequence of our choice we have $F_{\mu \nu}\equiv 0$. Then Eq.~(\ref{Fmunu}) gives immediately the relation between the Weyl vector and the scalar field as
\be
\omega ^\mu=\frac{1}{\alpha}\frac{\nabla ^{\mu}\Phi }{\Phi}.
\ee
Substituting this expression of the Weyl vector into the generalized Klein-Gordon equation (\ref{b5}0 it turns out that it is identically satisfied, $\Box \Phi-\Box \Phi\equiv 0$.  Explicitly, the relation between the scalar field and the Weyl vector is given by
\be\label{15}
\Phi '=\alpha \Phi \omega _1.
\ee

Next, we consider the gravitational field equations of the theory, which are given by two independent strongly nonlinear differential equations \cite{2,3}
\bea\label{c15}
&& -1+e^{\lambda }-\frac{1}{4}e^{\lambda }r^{2}\Phi-\frac{2r\Phi^{\prime }}{\Phi}+%
\frac{3r^{2}}{4}\frac{\Phi^{\prime 2}}{\Phi^{2}}+r\lambda ^{\prime }\left(
1+\frac{r}{2}\frac{\Phi^{\prime }}{\Phi }\right)\nonumber\\
&&-\frac{r^{2}\Phi^{\prime \prime
}}{\Phi}=0,
\eea
\bea \label{c16}
&& 1-e^{\lambda }+\frac{1}{4}e^{\lambda }r^{2}\Phi+\frac{2r\Phi^{\prime }}{\Phi}+\frac{%
3r^{2}}{4}\frac{\Phi^{\prime 2}}{\Phi^{2}}+r\nu ^{\prime }\left(1+\frac{r}{2}\frac{\Phi'}{\Phi}\right)
=0. \nonumber\\
\end{eqnarray}

A third field equation is a consequence of the above two equations. The system of Eqs.~(\ref{c15}) and (\ref{c16}) is not closed, since there are only two equations for three unknowns $\left(\nu,\lambda, \Phi\right)$. Hence, one more relation between the unknown variables is needed. By adding the field equations we obtain the evolution  equation for $\Phi$ as given by
\be\label{sum}
\frac{\Phi ''}{\Phi}-\frac{3}{2}\frac{\Phi ^{\prime 2}}{\Phi ^2}-\frac{\nu '+\lambda '}{r}\left(1+\frac{r}{2}\frac{\Phi '}{\Phi}\right)=0.
\ee

By imposing now the natural condition
\be
\nu +\lambda =0,
\ee
the evolution of the scalar field is decoupled from the evolution of the metric, and it is given by the equation
\be
\frac{\Phi ''}{\Phi}-\frac{3}{2}\frac{\Phi ^{\prime 2}}{\Phi ^2}=0,
\ee
with the general solution given by
\be\label{40}
\Phi(r)=\frac{C_1}{\left(r+C_2\right)^2},
\ee
where $C_1$ and $C_2$ are arbitrary integration constants. By denoting $u=re^{-\lambda}$, and by using the expression of $\Phi (r)$ as given by Eq.~(\ref{40}), from the field equations one obtains the differential equation
\bea\label{44}
&&\left(1-\frac{r}{C_2+r}\right)
   u'(r)+3\left[\frac{r}{(C_2+r)^2}-\frac{1}{C_2+r}\right] u(r)\nonumber\\
 &&  +\frac{C_1 r^2}{4 (C_2+r)^2}-1=0,
\eea
which has the solution \cite{2,3}
\bea\label{58}
re^{-\lambda}&=&\frac{r^2 \left(12 C_3 C_2^2+C_1-4\right)}{4 C_2}+r \left(3 C_3
   C_2^2+\frac{C_1}{4}-2\right)\nonumber\\
  && +C_3 C_2^3+\frac{1}{12}
   (C_1-12) C_2+C_3 r^3,
\eea
where by $C_3$ is an arbitrary integration constant. For a discussion of the various forms of the solution obtained by imposing different conditions on the integration constants see \cite{2}. Finally, for the radial component of the Weyl vector we obtain the expression
\be
\omega _1=\frac{1}{\alpha}\frac{\Phi '}{\Phi}=-\frac{2}{\alpha}\frac{1}{r+C_2},
\ee

\section{Discussions and final remarks}

In this short note we have presented a step by step derivation of an exact static spherically symmetric vacuum solution of the field equations of the Weyl geometric gravity theory. We have shown in detail that all the field equations are satisfied, and that the solution is correct mathematically. The solution is obtained under two basic assumptions: a) the Weyl vector has only a radial component, and b) the metric tensor components satisfy the relation $\nu +\lambda=0$. With respect to point b), we would like to mention that most of the vacuum solutions in general relativity, or modified gravity theories, satisfy this condition. The choice of the Weyl vector gives the condition $F_{\mu \nu}\equiv 0$, and thus the Weyl vector is given as the gradient of the logarithm of the scalar field,
\be
\omega _1=\frac{1}{\alpha} \frac{d}{dr}\ln \Phi.
\ee
This shows that the solution belongs to the particular class of Weyl geometry, called the Weyl Integrable Geometry.   In the Weyl Integrable Geometry the connection differs from the Levi-Civita connection by a scalar field related to a  conformal metric transformation. This is a valid mathematical structure, which has been investigated in many physical contexts \cite{7,8}, and there is no a priori reason for claiming that such a solution cannot describe, for example, the galactic rotation curves, as suggested in \cite{2}. We would like to point out that except the above mentioned conditions a) and b) no other conditions have been used in the derivation of the solution, which is obtained strictly from the three main equations of motion for the metric tensor, the scalar field, and the Weyl vector. In particular, no gauge condition of the form $\nabla _\mu \omega ^\mu =0$ has been imposed on the Weyl vector, as wrongly claimed in \cite{1}. But still there is a definite relation between the Weyl vector and the scalar field, which follows naturally from  the field equations, and which can be interpreted as a gauge fixing procedure.

To conclude: the exact solution found in \cite{2} is correct mathematically, and it satisfies all the equations of motion of the theory. Of course this is a simple particular solution of a very rich theoretical model, which may have several other solutions, exact or numerical. However, the investigations carried out in \cite{2} indicate that despite its simplicity, the exact solution of the Weyl geometric gravity has the potential of giving important insights into some of the fundamental problems of the present day astrophysics and cosmology. In this sense it is a first step towards the extensive and full investigations of the physical implications of the Weyl geometric structures, and of the gravitational theories based on them.

\section*{Acknowledgments}

The work of TH is supported by a grant of the Romanian Ministry of Education and Research, CNCS-UEFISCDI, project number PN-III-P4-ID-PCE-2020-2255 (PNCDI III).

\end{document}